\documentclass[twocolumn,english,aps,superscriptaddress,prb]{revtex4-2}

\usepackage[colorlinks=true,allcolors=blue]{hyperref}
\usepackage{array}
\usepackage[latin9]{inputenc}
\usepackage{amsmath}
\usepackage{amssymb}
\usepackage{graphicx}
\usepackage{babel}
\usepackage{mathrsfs}
\usepackage{amsfonts}
\usepackage{epstopdf}
\usepackage{multirow}
\usepackage{color}
\usepackage{natbib}
\usepackage{bm}

\usepackage{xcolor}

\begin{document}

\title{Holographic Representations of Topological Quantum Criticality:\\ Emergent Symmetry Approach around the Bott Clock}

\author{Fan Yang}
\affiliation{Key Laboratory of Quantum Materials and Devices of Ministry of Education, School of Physics, Southeast University, No.2 SEU Road, Nanjing, China 211189}

\author{Fei Zhou}
\affiliation{Department of Physics and Astronomy, University of British Columbia, 6224 Agricultural Road, Vancouver, BC, Canada V6T 1Z1}

\begin{abstract}
In this article, we apply the idea of {\em emergent symmetries} to construct holographic representations of a broad class of topological quantum critical points (tQCPs) that appear in fermionic classes in the  Bott clock. We show explicitly that around the Bott clock, an emergent symmetry $U_{EM}$ exists at a tQCP between different symmetry protected gapped phases with protecting symmetry $G_p$ in $d$ spatial dimensions. This emergent symmetry can be used to construct a $(d+1)$ spatial dimensional lattice model with a properly upgraded symmetry such as $U_{EM}\times G_p$ or $U_{EM} \rtimes G_p$, etc. By doubling the degrees of freedom of the adjacent topological class in $d$-dimensions, we successfully show that the $(d+1)$-dimensional lattice models constructed in this way have the desired enlarged symmetry groups that precisely belong to the adjacent topological classes, counterclockwise around the Bott clock. 
Furthermore, $d$-dimensional boundaries of these $(d+1)$ dimensional lattice models of gapped topological classes are shown to exhibit identical infrared dynamics as the corresponding tQCPs in the $d$-dimensional adjacent topological phases in the Bott clock, with lower symmetries.
\end{abstract}

\maketitle
\section{Introduction}
Numerous recent efforts to understand topological quantum critical points (tQCPs) indicate a few surprising features \cite{Tsui15,Jiang18,Verresen18,Bi19,Tsui19,Huang20,Zhou22,Zhou23,Zhou24}.
One of the most fascinating of them is perhaps the emergent symmetries and
symmetry anomalies associated with them \cite{Tsui19,Huang20,Zhou22,Zhou23,Zhou24}.
Although it is not entirely clear how different studies and observations in a wide variety of topological matters are connected,
there appear to be two major trains of thought that have been pursued.

One is very much related to applications of generalized symmetries in even conventional order-disorder phase transitions \cite{Lake18,Zhao21}, where the symmetry concept has been generalized to include dual symmetries, mostly of higher forms in high dimensions and topological theories \cite{Ji20,Chatterjee21}. This opens up a new 
frontier in theoretical studies of quantum criticality via higher-form extended topological excitations. These studies further shed new light on symmetry enriched topological orders in general \cite{Wen02,Essin12,Mesaros13,Lu16,Barkeshili19}.

The other big class of states that have recently received intense attention are the fermionic topological states, either invertible topological states such as symmetry protected topological (SPT) states \cite{Hasan10,Qi11,Bernevig,Schnyder08,Kitaev09,Chiu16,Wen12,Senthil15} or more exotic noninvertible states \cite{Seifnashiri24,Meng24,Aksoy25}. Here new symmetries or invariance can emerge in the infrared (IR) at tQCPs in topological states where at a more fundamental level or in generic gapped lattice models, these symmetries have to be broken in order to define such topological order \cite{Tsui19,Huang20,Zhou22,Zhou23,Zhou24}.
These emergent symmetries are very surprising, and their detailed mechanisms or systematics are not understood.

One well known example is the emergent time reversal symmetry (TRS) at a tQCP in TR breaking $p+ip$ D class superconductors \cite{Read00,Tsui19}.
These are topological states that can be defined only when TRS is broken and are protected only by the charge-conjugation symmetry.
Nevertheless, at the tQCP, the IR effective field theory (EFT) does have an emergent TRS.

The EFT does capture a single Majorana cone dynamics in the IR with a central charge equal to $c=\frac{1}{4} \mod (\frac{1}{2})$.
Such a small degree of freedom is forbidden in an UV completed theory or in a 2D bulk lattice. In addition, a single {\em gapless cone} at a tQCP is also strictly forbidden if TRS is an on-site symmetry. Both imply the anomalous nature of such low energy dynamics and the emergent symmetry accompanying them.

In fact, technically such an EFT can be closely related to the surface theory of a 3D topological superconductors where TRS becomes anomalous near a surface because of the bulk inflow. This technical similarity suggests a holographic nature of the 2D tQCP from the point of view of the emergent symmetry, although it is not fully understood why only TRS emerges at this tQCP, but not other symmetries.

Another less obvious example is a tQCP in 3D DIII class topological superconductors with protecting symmetry $G_p=Z^T_2$.
In this case, one can show that the tQCP IR dynamics can be mapped into a single Weyl cone dynamics with an emergent $U(1)$ symmetry (of chiral nature) \cite{Zhou22,Zhou23,Zhou26}. A single Weyl cone is strictly forbidden in a 3D lattice with an on-site charge $U(1)$ symmetry because of the celebrated no-go theorem on the fermion doubling problem \cite{Nielsen81}.

In this case, one can argue that the emergent symmetry group is naturally related to possible continuous deformation around the point of tQCP in the surrounding gapped states \cite{Zhou24}.
A tQCP always separates two distinct SPT phases under the protection of symmetry $G_p$. The deformation, on the other hand, when allowed to break the protecting symmetry $G_p=Z^T_2$, can smoothly connect two topologically distinct SPT phases. So in this particular case, the mechanism appears to be much better understood---as the deformation group, which is local unitary, can leave the tQCP invariant if the tQCP is a typical isolated conformal field theory. 

Again in this case, one can explicitly construct a ($4+1$)D topological insulator with $G_p = U(1) \rtimes Z_2^T$ with $U(1)$ matching the emergent symmetry at the $3$D tQCP and show that the surface can be identical to the tQCP in the IR limit and supports a single Weyl cone at a boundary.

The boundaries or surfaces of SPT states are usually understood as a result of symmetry fractionalization, that the on-site protecting symmetry manifests itself in a non-local way on boundaries, therefore demanding gapless dynamics. These symmetry anomalies are broad manifestation of 't Hooft anomalies \cite{tHooft74} and have been widely applied to detect gapless edges or surfaces in SPTs. They become quite well-understood.

The emergent symmetry at tQCPs on the other hand is a much trickier issue in general. This is because such symmetries are not in the original lattice model or any UV completed models.
Rather, they are unique features of IR fixed points. 
Near an IR fixed points, a simple EFT that captures IR dynamics often seems to exhibit additional symmetries not generic in their UV completed lattice models.

At the moment, for fermion topological states we are not aware of a systematic way to derive these emergent symmetries from the protecting symmetry group $G_p$ and topological data such as boundary states.
The general intuitions based on previous studies are these emergent symmetries $U_{EM}$ (the subscript refers to the {\em emergent symmetry}), being anomalous, can also be, at least mathematically, characterized as a symmetry with 't Hooft anomalies. This is analogous to the 't Hooft anomalies of protecting symmetries at gapless boundaries of a symmetry protected states, although here $U_{EM}$ is not an exact global symmetry to start with. 

Technically, there is a procedure to differentiate an emergent symmetry appearing in {\em local} IR dynamics from an exact {\em global} symmetry in a complete UV model.
It can be carried out by adding a global or topological term breaking the emergent symmetry. It effectively explicitly marks two distinct 
symmetries at the level of EFTs: an anomalous emergent symmetry, $U_{EM}$ in local IR dynamics vs an exact global symmetry or the protecting symmetry $G_p$ in UV completed dynamics in our specific context. 't Hooft anomalies can be applied for our purpose of labeling $U_{EM}$. Such a framework establishes a natural connection between an emergent anomalous symmetry at a $d$-dimension tQCP, and 't Hooft anomalies at the boundaries of a $(d+1)$-dimensional topological state. Instead of a bulk inflow of symmetry anomalies to boundaries of a topological bulk, here we have a flow of emergent symmetry anomalies towards infrared from a UV completed model. 

In this article, we simplify the procedure by directly matching the anomalous degrees of freedom at a tQCP implied by emergent symmetries with the surface gapless fermions in one-dimension higher. The matching conditions
consistent with the protecting symmetry $G_p$ and emergent symmetries $U_{EM}$ at a tQCP naturally point to a $(d+1)$-dimensional topological bulk with upgraded protecting symmetry $\tilde{G}_p$. This is a practical way of matching 't Hooft anomalies in two sets of models.

The gapless degrees of freedom $N_f$ at tQCPs in SPTs can be uniquely defined by two sets of data:

\begin{itemize}
\item the change of gapless surfaces or boundary modes across the tQCP, $\delta \nu$;

\item the minimum degrees of freedom or the flavor of fermions, $N^0_f (d,G_p)$ in the fundamental representation of the protecting symmetry $G_p$ in $d$ dimensions. As expected, $\delta \nu=1$ in such a representation. (Here, a pair of boundary modes protected by symmetry counts as one) 
\end{itemize}

Given these considerations, we can determine that
\begin{equation}
N_f(d, G_p;\delta \nu)=N_f^0(d, G_p) \delta \nu, \quad\delta \nu=1,2,3,... 
\label{DF}
\end{equation}
This relation, which bypasses the changes of topological invariants across the transition, is a result of two topological aspects:

(a) An index theorem which relates the gapless modes to the topological invariants; i.e.
\begin{equation}
N_f=N_f^0(d, G_p) \frac{\delta N_w}{\delta N^0_w}, \quad\delta N_w=1,2,3,... 
\label{index}
\end{equation}
where $\delta N^0_w (d, G_p)$ is the minimum change of the topological invariants consistent with the protecting symmetry $G_p$.

(b) A holographic principle which relates a topological invariant ${N_w}/{\delta N^0_w(d,G_p)}$ to its 
boundaries with gapless modes specified by $\nu=1,2,3,...$; i.e.
\begin{equation}
\frac{\delta N_w}{\delta N^0_w}=\delta \nu,  \quad
\delta \nu=1,2,3,... .
\label{holograph}
\end{equation}

In 3D DIII class states, we have $N_f^0(Z_2^T)=1/2$, or one half of a Dirac fermion in 3D. So when $\delta \nu=1,3,5...$, tQCPs can have odd numbers of Weyl cones which evade the standard lattice chiral fermion no-go theorem with an exact charge $U(1)$ symmetry.

In D class topological states in $2D$, we have $N_f^0=1/2$, one half of 2D Dirac fermions, or one Majorana fermion. For $\delta \nu=1,3,5,...$, this can lead to odd numbers of gapless Majorana cones at tQCPs with an emergent TRS. These appear to be anomalous structures for 2D bulk lattices with TRS.

This flavor-boundary relation in Eq. (\ref{DF}) basically dictates the infrared dynamics (i.e. $N_f$) and emergent symmetries at tQCPs via the topological data on the boundary of gapped phases. In (2+1)D, it also leads to the correct counting of central charge $c$ appearing in topological orders that can appear in these symmetry protected states.

Therefore, in this article instead of researching the mechanism of why an emergent symmetry appears at a tQCP with protecting symmetry $G_p$,
we focus on the following explicit holographic relations in the tenfold way classification of fermion topological states via the Bott clock structure.

The key relation to establish is between

(a) a tQCP in the $d$-dimensional lattice with a protecting symmetry $G_p$, which further exhibits an emergent symmetry $U_{EM}$ in the EFT (the emergent symmetry can be either unitary or antiunitary); and

(b) the $d$-dimensional surface of a $(d+1)$-dimensional SPT with the enlarged protecting symmetry $\tilde G_p$ with $G_p$ and $U_{EM}$ as its subgroups, such as $\tilde{G}_p=G_p \times U_{EM}$ or $G_p \rtimes U_{EM} \mbox{  or  }  G_p \ltimes U_{EM}$, ect.

We will present the explicit relations in the Bott clock of the fermion SPTs (Fig. \ref{fig:bott}). Going around the clock counterclockwise, we are able to construct a holographic representation of a $d$-dimension tQCP with an emergent symmetry, utilizing the boundary of a one-dimension higher gapped SPT with higher protecting symmetries which is the subsequent class of the $d$-dimensional gapped SPT in the Bott clock . 

\begin{figure}
    \centering
    \includegraphics[width=\columnwidth]{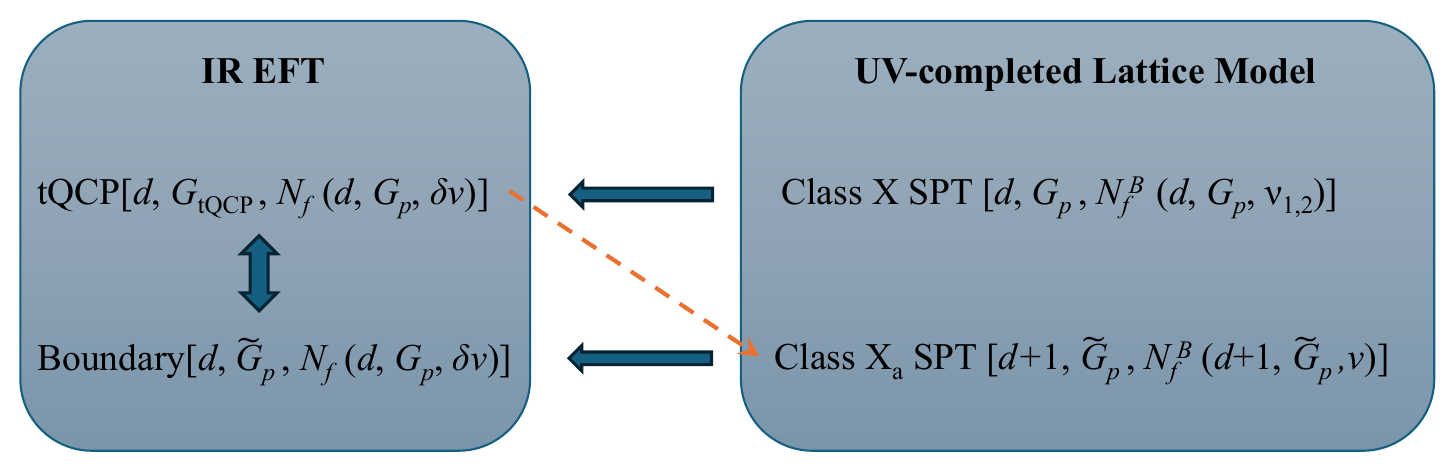}
    \caption{Procedure of constructing the holographic theory of a tQCP. The left panel shows the IR EFTs and the right panel shows the UV-completed lattice models. We start with a $d$-dimensional lattice model of class-X with protecting symmetry $G_p$ and bulk lattice degrees of freedom $N^B_f(d, G_p, \nu_{1,2})$ (the superscript refers to the bulk) where $\nu_1$ and $\nu_2$ are the numbers of gapless boundary states of two gapped topological phases respectively. The IR EFT of the tQCP in the class X has an enlarged symmetry $G_\text{tQCP}=G_p\times U_{EM}$, where $U_{EM}$ is an emergent symmetry, and gapless fermions $N_f(d, G_p, \delta \nu)$. (Note the group structure may also be more complicated than a direct product depending on details of the model.) The data of $G_\text{tQCP}$ and $N_f$ can be applied to construct a $(d+1)$-dimensional gapped lattice model of class-$X_a$ which has an enlarged protecting symmetry $\tilde G_p=G_\text{tQCP}$, with double the fermion degrees of freedom, i.e. $N_f^B(d+1,\tilde G_p, \nu)=2 N_f(d, G_p, \delta \nu)$ (see Eq\ref{DF})\footnote{We assume $N^B_f(d, G_p, \nu_{1,2})$
    is large enough to support $\nu_{1,2}$ gapless surface states. For all explicit constructions below, we will simply set $\delta \nu=1$. See discussions in the main text.}. The tQCP has the same IR dynamics as a $d$-dimensional boundary state with protecting symmetry $\tilde G_p=G_\text{tQCP}$. This establishes the holographic relation between the $d$-dimension tQCP in class-X with protecting symmetry $G_p$ and the $(d+1)$-dimensional gapped SPT with protecting symmetry $\tilde{G}_p$ in one higher dimension in the adjacent class-$X_a$ in the Bott clock.}
    
    \label{fig:diagram}
\end{figure}

In exploring and constructing these holographic theories of tQCPs, we follow the following blue-print closely (See also Fig. \ref{fig:diagram}):

(0) We start with a gapped $d$-dimensional topological state in class-$X$ in the Bott clock in a UV completed lattice model. This lattice model can be equipped with a fundamental representation of a protecting symmetry $G_p$ for class-$X$ with fermion flavor given as $N^0_f(d,G_p)$. We consider a transition between two gapped topologically distinct states with the number of gapless boundary modes given by $\nu_{1}$ and $\nu_2$, respectively.

(I) We first identify an IR emergent symmetry $U_{EM}$ at tQCPs, 
which separate two topologically distinct gapped phases 
in the $d$-dimensional UV completed lattice,
specified by $\nu_{1}$ and $\nu_2$ gapless boundary states in Class-$X$ with a protecting symmetry $G_p$. The overall symmetry $G_\text{tQCP}$ in the IR limit
can be $U_{EM} \rtimes G_p$, $U_{EM} \ltimes G_P$, $U_{EM} \times G_P$, or more complicated structure, 
depending on details of $G_p$ and $U_{EM}$ (See explicit constructions in Sec.\ref{sec:real}). The fermion flavor is specified in Eq. (\ref{DF}).

(II) We then upgrade the protecting symmetry group from $G_p$ to $\tilde{G}_p=G_\text{tQCP}$ by doubling the fermion degrees of freedom in the fundamental representation of the tQCPs to $N_f^0(d+1,\tilde{G}_p)=2 N^0_f(d,G_p)$. 
Simultaneously, we also elevate the theory to the $(d+1)$-dimensional lattice model by adding an additional dimension to the original $d$-dimensional lattice model. 

(III) A $(d+1)$-dimension lattice model constructed above always turn out to belong to class $X_a$ exactly adjacent to class $X$ in the Bott clock along the counterclockwise direction. Its gapped topological phases are in Class $X_a$ (the subscript $a$ refers to adjacent) with protecting symmetry $\tilde{G}_p$.
A specific gapped phase is further selected by matching the number of gapless boundary states with $N_f(d,G_p,\delta \nu)$ in Eq. (\ref{DF}).

\begin{figure}
    \centering
    \includegraphics[width=\columnwidth]{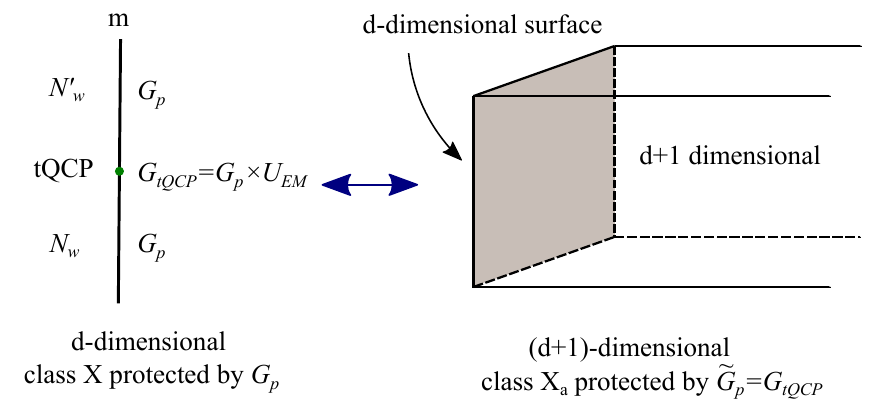}
    \caption{Left: Fermionic SPTs of class X in $d$-dimensions protected by symmetry group $G_p$. A tQCP separates two topological phases of topological invariants $N_w$ and $N'_w$ when $m$ is tuned. At tQCP, there is an emergent symmetry $U_{EM}$ in the infrared EFT, giving $ G_\text{tQCP}=G_p\times U_{EM}$ (or more complicated group structure depending on details.) Right: A $(d+1)$-dimensional gapped SPT of class $X_a$ with protecting symmetry $\tilde G_p=G_\text{tQCP}$, whose $d$-dimensinoal surface has the same IR dynamics as the $d$-dimensional tQCP of class X. This suggests a holographic theory of tQCP.}
    \label{fig:illustration}
\end{figure}

(IV) The $d$-dimensional gapless boundaries of the chosen topological phases in class $X_a$ with upgraded protecting symmetry $\tilde{G}_p$ exhibit identical 
infrared dynamics as the $d$-dimensional tQCPs in class-$X$ in the Bott clock with the protecting symmetry $G_p$. The $d$-dimensional boundaries obtained this way therefore form a holographic representations of tQCPs in the $d$-dimensional bulk with $G_p$. See also Fig. \ref{fig:illustration}

For tQCPs in 2D D class and 3D DIII class, explicit lattice model constructions were put forward previously in Ref. \cite{Tsui19} and Ref. \cite{Zhou24}, respectively. We refer readers to these articles for details.

In this article, we show it is possible to construct similar holographic theories for all tQCPs and they can be organized systematically around the Bott clock. And practically without loss of generality, we restrict ourselves to the cases of $\delta \nu=1$, i.e. the fundamental representations.

The article is organized as follows.
In Sec. \ref{sec:symmetry}, we review the Bott clock of topological insulators and superconductors and the microscopic symmetry realization of each class. In Sec. \ref{sec:holography}, we summarize the general framework of the holographic theory of tQCPs on a hypercubic lattice. In Sec. \ref{sec:real} and Sec. \ref{sec:complex}, we give explicit examples of the holographic theory for each symmetry class. Finally, we conclude this article with remarks and open question in Sec. \ref{sec:discussion}.

\section{Bott clock and symmetry realization in free fermion systems}\label{sec:symmetry}

Before presenting our results, we first review the tenfold way classification of free fermion systems and the microscopic symmetry realization of each class.

The single-particle Hamiltonian matrix $H$ of free fermion systems can be classified by three symmetries, namely, time-reversal (TRS), particle-hole (PHS), and sublattice symmetries. In the matrix representation, time-reversal and particle-hole symmetries are antiunitary, which can be either absent (denoted as 0), or squared to $+1$ or $-1$. Their product is the sublattice symmetry, which is unitary and anticommutative with $H$. The presence of sublattice symmetry is determined by the other two antiunitary symmetries, except when both TRS and PHS are absent, in which case sublattice symmetry can be either absent or present. As a result, there are in total $3\times 3+1=10$ symmetry classes, commonly referred to as the Atland-Zirnbauer (AZ) classes \cite{Zirnbauer96,Altland97,Schnyder08}.

\begin{figure}
    \centering
    \includegraphics[width=0.75\columnwidth]{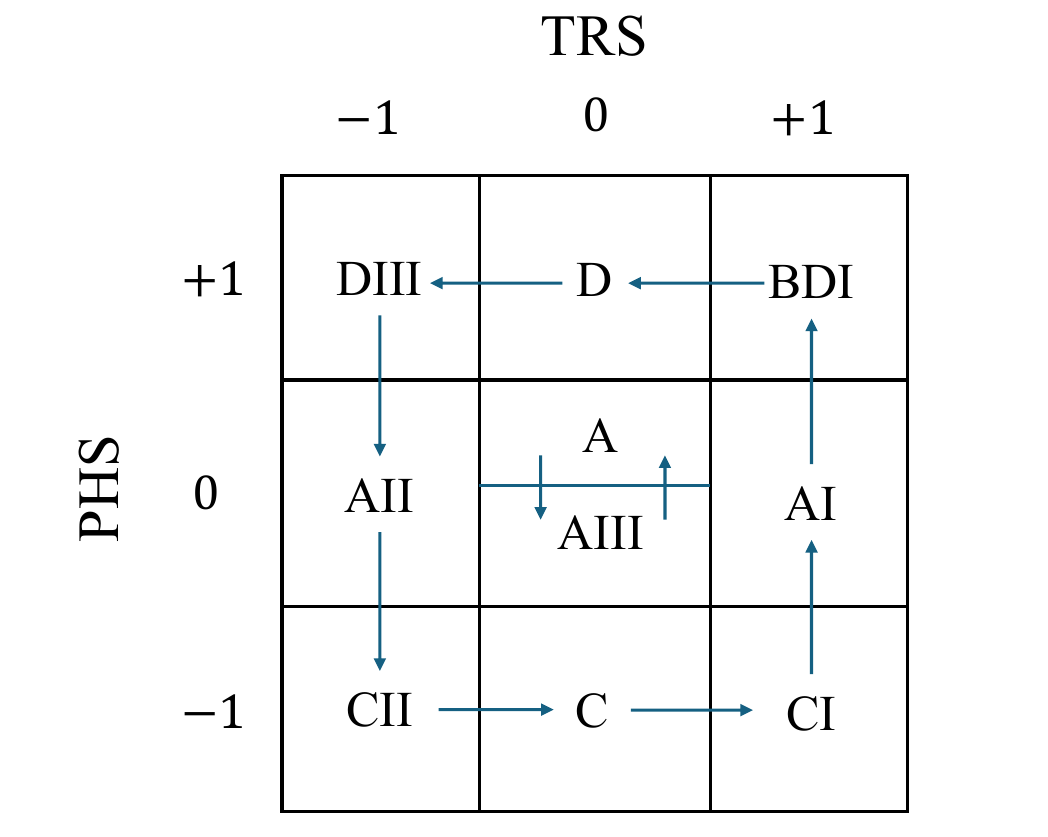}
    \caption{Bott clock of topological insulators and superconductors. $+1$ and $-1$ represents the square of TRS and PHS when they are present, and $0$ means the absence of the symmetry. Following the direction of the arrows, the topological classification remains the same when spatial dimension is increased by one.}
    \label{fig:bott}
\end{figure}

These ten symmetry classes can be divided into two sets, one comprising two classes A and AIII, and the other comprising the remaining eight classes D, DIII, AII, CII, C, CI, AI, and BDI. By arranging each symmetry class in this order within each set, a periodic table emerges. When moving from one class to the next and simultaneously raising the spatial dimension by one, the topological classification of the Hamiltonian remains the same. This is the famous periodic table of topological insulators and superconductors discovered by Kitaev \cite{Kitaev09}. 
This classification follows the Bott periodicity, and therefore, these ten classes can be organized on a so-called Bott clock shown in Fig. \ref{fig:bott}. The two classes in the center are referred to as complex classes and the remaining eight classes are referred to as real classes. Following the arrows within each sequence, the topological classification remains the same when the spatial dimension is also increased by one.

It is worth noting that in this tenfold way classification, the symmetries of the Hamiltonian matrix, i.e., TRS, PHS, and sublattice symmetry, may not be the microscopic symmetry of the fermionic system. For example, since fermions have half-integer spins, the microscopic TRS always squares to $-1$. However, if SU(2) spin rotation symmetry is imposed, the system will have an apparent TRS squaring to $+1$. On the other hand, within the meanfield description, superconductors always have an emergent PHS squaring to $+1$ due to the particle-hole redundancy of Nambu space. Microscopically, this emergent PHS is resulted from the breaking of charge conservation and reflects fermion parity conservation. In the following, we will focus on the microscopic symmetry realization of each AZ class with  spin-$1/2$ fermions. We will also assume translation symmetry and write both the second quantized Hamiltonian $\mathcal{H}$ and its matrix representation $H$ in the momentum space.

For the eight real classes, we always write the Hamiltonian in terms of Majorana fermions
\begin{equation}\label{Majorana}
    \mathcal{H}=\sum_{\bf k} \chi_{\bf -k}^T H({\bf k})\chi_{\bf k},
\end{equation}
where $\chi=(\chi_{1+},...,\chi_{n+},\chi_{1-},..,\chi_{n-})^T$ is a $2n$-component Majorana fermion.
In the Majorana basis, the Hamiltonian matrix $H({\bf k})$ always has the following symmetry
\begin{equation}\label{FermiConstraint}
    KH({\bf k})K^{-1}=-H(-{\bf k}),
\end{equation}
with $K$ being complex conjugation, as a result of anticommutation relation between real Majorana fermions. In the tenfold way classification, this can be interpreted as a PHS with $K^2=1$. But microscopically, this is simply a consequence of the particle-hole redundancy in the Nambu representation. Following Refs. \cite{Wen12,Kennedy16}, we will refer to Eq. (\ref{FermiConstraint}) as the Fermi constraint and not consider it as a symmetry. 
The Fermi constraint dictates that all the terms odd in ${\bf k}$ must couple to real symmetric matrices and all the terms even in ${\bf k}$ must couple to imaginary antisymmetric matrices in the Majorana representation.

Aside from the Fermi constraint, without imposing any symmetry, we obtain class D superconductors.
If we impose the microscopic TRS of half-integer spins
\begin{equation}\label{TRS}
    TH({\bf k})T^{-1}=H(-{\bf k}),\quad T^2=-1,
\end{equation}
we obtain class DIII. 
We note that in the Majorana basis, the unitary part $U_T$ of TRS $T=U_TK$ always anticommute with $H({\bf k})$ due to the Fermi constraint (\ref{FermiConstraint}), i.e.,
\begin{equation}
    U_TH({\bf k})U_T^{-1}=-H({\bf k}).
\end{equation}
In addition to $T$, if we further impose charge conservation $Q$, we obtain an insulator with $T^2=-1$, which is the symplectic class AII. On top of AII, if we impose PHS
\begin{equation}\label{PHS}
    CH({\bf k})C^{-1}=-H(-{\bf k}),\quad C^2=-1,
\end{equation}
its combination with TRS lead to a sublattice symmetry, resulting in the chiral symplectic class CII.
Due to Fermi constraint, in the Majorana basis the unitary part $U_C$ of PHS $C=U_CK$ commutes with $H({\bf k})$
\begin{equation}
    U_CH({\bf k})U_C^{-1}=H({\bf k}).
\end{equation}

On the other hand, if we start from class D and impose spin SU(2) symmetry, we obtain a spin-singlet superconductor. The spin SU(2) symmetry combined with the Fermi constraint $K$ results in an apparent PHS squared to $-1$, and the superconductor belongs to class C.
If we further impose microscopic TRS $T^2=-1$, its combination with SU(2) symmetry appears as a TRS for spinless particles, thus realizing class CI.
On top of CI, if we impose charge conservation $Q$, we reach an insulator with an apparent TRS for spinless particles, which belongs to orthogonal class AI. If we impose PHS $C^2=-1$ on AI, its combination with SU(2) appears as a PHS squared to $+1$, resulting in the chiral orthogonal class BDI. We emphasize that the SU(2) generators are all imaginary in the Majorana basis, and the SU(2) rotations must commute with $T$, $C$, and charge U(1) symmetry $U=e^{i\theta Q}$. The existence of SU(2) effectively flips the sign of second and third column for the four class with SU(2) symmetry (see Table \ref{tab:Symmetry}).

For the two complex classes, we first write the many-body Hamiltonian in terms of complex fermions 
\begin{equation}\label{ComplexClass}
    \mathcal{H}=\sum_{\bf k}\psi_{\bf k}^\dagger H({\bf k}) \psi_{\bf k} 
\end{equation}
where $\psi=(\psi_1,\psi_2,...,\psi_n)^T$ is an $n$-component complex fermion. 
Eq. (\ref{ComplexClass}) always possesses a charge conservation symmetry $Q$. Without imposing any other symmetry, $H({\bf k})$ belongs to the unitary class A. If we impose sublattice symmetry $S$
\begin{equation}\label{CS}
    SH({\bf k})S^{-1}=-H({\bf k}),
\end{equation}
we obtain the chiral unitary class AIII.

The sublattice symmetry $S$ is often referred to as a pseduosymmetry as it anticommutes with the Hamiltonian matrix $H({\bf k})$. 
So when discussing microscopic symmetries, it is beneficial to use Majorana basis Eq. (\ref{Majorana}). In this case, we have $\psi_i=(\chi_{i+}+i\chi_{i-})/\sqrt{2}$.
For class A, the charge conservation $Q$ in this Majorana representation becomes a pseduospin U(1) symmetry $U=e^{i\theta s_y}$, where $s_y$ is the purely imaginary antisymmetric Pauli matrix in the $\{+,-\}$ subspace. 
The sublattice symmetry $S$ is equivalent to a TRS $T=SK$ in the Majorana basis as $S$ anticommutes with $H({\bf k})$.  Therefore, class AIII can also be realized by a spin U(1) symmetry and a TRS.

We summarize the above microscopic symmetry realization of each AZ class in Table \ref{tab:Symmetry}. We note that the symmetry realization of each AZ class is not unique. Some examples were discussed in more details in Ref. \cite{Schnyder08}. In this paper, the symmetry realization in Table \ref{tab:Symmetry} closely follows Ref. \cite{Kennedy16}. A more comprehensive analysis on symmetry realization based on TRS, PHS, and charge $Q$ can be found in Ref. \cite{Wen12}.
In the following sections, we show that using the symmetry realization in Table \ref{tab:Symmetry}, one can systematically construct the emergent symmetry at tQCPs and the holographic representation of these tQCPs.

\begin{table}[]
    \centering
    \begin{tabular}{|c|c|c|c|l|}
    \hline
    class  & TRS & PHS & Chiral & symmetry realization \\
    \hline
    A & 0 & 0 & 0 & $Q$ or Spin U(1)\\
    \hline
    AIII & 0 & 0 & 1 & $Q$, $S$ or Spin U(1), $T$\\
    \hline
    D & 0 & 1 & 0 & none\\
    \hline
    DIII & -1 & 1 & 1 & $T$\\
    \hline
    AII & -1 & 0 & 0 & $T$, $Q$\\
    \hline
    CII & -1 & -1 & 1 & $T$, $Q$, $C$\\
    \hline
    C & 0 & -1 & 0 & SU(2)\\
    \hline
    CI & 1 & -1 & 1 & SU(2), $T$\\
    \hline
    AI & 1 & 0 & 0 & SU(2), $T$, $Q$\\
    \hline
    BDI & 1 & 1 & 1 & SU(2), $T$, $Q$, $C$\\
    \hline
    \end{tabular}
    \caption{Microscopic symmetry realization of the 10 AZ classes of spin-$1/2$ fermions. Column 2 to 4 list the symmetries of the Hamiltonian matrix $H$ in the tenfold way classification. $\pm 1$ denotes the square of each symmetry when it is present. $0$ denotes the absence of this symmetry. The last column is the microscopic symmetry realization of each class. $S$ is the unitary sublattice symmetry defined in Eq. (\ref{CS}). $Q$ is charge conservation. $T$ and $C$ are antiunitary TRS and PHS defined in Eqs. (\ref{TRS},\ref{PHS}) with $T^2=-1$ and $C^2=-1$. SU(2) is the spin rotation symmetry. We note that for each AZ class, there can be multiple microscopic symmetry realizations. The last column can be viewed as a sufficient but not necessary condition. }. 
    \label{tab:Symmetry}
\end{table}

\section{Holographic theory of fermionic quantum critical points}\label{sec:holography}
In this section, we give the generic form of the $(d+1)$-dimensional bulk lattice model of class $X_a$ whose boundary has the same IR dynamics as a tQCP of class $X$. In the two following sections, we give the explicit construction for each symmetry class. Our construction is a generalization of Ref. \cite{Tsui19} and Ref. \cite{Zhou24}.

Let us consider free fermion models of class $X$ on $d$-dimensional hypercubic lattices with symmetry $G_p$ \cite{Tsui19}
\begin{equation}\label{latticeH}
 H({\bf k})=\sum_{i=1}^d\Gamma_i \sin k_i +[m+d-\sum_{i=1}^d\cos k_i]\Gamma_0,
\end{equation}
where 
$\{\Gamma_i,\Gamma_j\}=\delta_{ij}$. When the Hamiltonian is written in the basis of $2n$-component Majorana fermions $\chi_{\bf k}$, $\mathcal{H}=\sum_{\bf k} \chi_{-\bf k}^T H(k)\chi_{\bf k}$, the real anticommutation relation of Majorana fermions further requires $\Gamma_i, i=1,2,...,d$ to be real symmetric, and $\Gamma_0$  imaginary antisymmetric.
When the Hamiltonian is written in the basis of $n$-component complex fermions $\psi_{\bf k}$,  $\mathcal{H}=\sum_{\bf k}\psi_k^\dagger H({\bf k})\psi_{\bf k}$, we only require $\Gamma_i, d=0,1,...,d$ to be Hermitian.
In the following, we focus on the tQCP at $m=0$, which separates two topologically distinct phases at $m>0$ and $-2<m<0$. Here, $m'=d-\sum_{i=1}^d\cos k_i$ is the the regularization mass that guarantees a single gapless point in the Brillouin zone at $\bf k=0$ for $m=0$. We have a trivial phase for $m>0$ and a topological phase for $-2<m<0$ \cite{Tsui19}.

The low-energy EFT of the lattice model has the Dirac form
\begin{equation}
    H(k)=\sum_{i=1}^d\Gamma_i k_i +m\Gamma_0,
\end{equation}
Such Dirac EFTs can describe the low energy physics of  topological phase transitions at $m=0$ and yield the correct change of topological invariants across the tQCPs.
The continuum EFT acquires an emergent symmetry $U_{EM}$ at the tQCP, and the symmetry at the tQCP becomes $G_\text{tQCP}=G_p\times U_{EM}$ or some more complicated group structure (see the two following sections). This enables us to connect the $d$-dimensional tQCP and the boundary of a $(d+1)$-dimensional bulk belonging to the subsequent class in the Bott clock, and the $(d+1)$ dimensional bulk Hamiltonian should have double the fermion degrees of freedom. It is important to note that the emergent symmetries of the tQCP only exist in the EFT. In the lattice model Eq. (\ref{latticeH}), the regularization mass $m'$ explicitly breaks the emergent symmetries at $m=0$ for ${\bf k}\neq0$, reflecting anomalies of the emergent symmetry \cite{Tsui19,Huang20}.
For class A, the emergent symmetry is the sublattice symmetry $S$ or a TRS that commutes with U(1). For real classes, the emergent symmetry is the additional symmetry of the subsequent class (see the last column of Table \ref{tab:Symmetry}), except for class CII, whose emergent symmetries and connection to class C will be discussed in detail in Sec. \ref{sec:SU2}. The explicit construction of emergent symmetries are given in Sec. \ref{sec:real} and Sec. \ref{sec:complex} for each symmetry class.

The symmetry group $G_\text{tQCP}$ is the same as the protecting symmetry of the subsequent class $X_a$ in the Bott clock. Therefore, we can construct a $(d+1)$-dimensional bulk EFT with $\tilde G_p=G_\text{tQCP}$ as
\begin{equation}
    \tilde{H}({\bf k})=\sum_{i=1}^d \gamma_z\Gamma_ik_i +\gamma_x k_{d+1}+M\gamma_y,
\end{equation}
where $\gamma_{x,y,z}$ are Pauli matrices. 
We can regularize this EFT on a hypercubic lattice and obtain the $(d+1)$-dimensional lattice Hamiltonian of class $X_a$,
\begin{equation}
\begin{split}
    \tilde{H}({\bf k})=&\sum_{i=1}^d \gamma_z\Gamma_i\sin k_i +\gamma_x \sin k_{d+1}\\
    &+[M+d+1-\sum_{i=1}^{d+1}\cos k_i]\gamma_y,
\end{split}
\end{equation}
which is topological when $-2<M<0$ \cite{Tsui19}.
The $(d+1)$-dimensional bulk theory has double the fermion degrees of freedom of the $d$-dimensinoal tQCP.
In the complex fermion representation, $\tilde{\mathcal{H}}=\sum_{\bf k}\Psi_k^\dagger \tilde H({\bf k})\Psi_{\bf k}$, with $\Psi_{\bf k}$ a $2n$-component complex fermion, $\psi_{\bf k}=P\Psi_{\bf k}$, and $P=(1+\gamma_z)/2$; in the Majorana representation $\tilde{\mathcal{H}}=\sum_{\bf k} X_{-\bf k}^T \tilde H({\bf k}) X_{\bf k}$, with $X_{\bf k}$ a $4n$-component Majorana fermion and $\chi_{\bf k}=PX_{\bf k}$.
The matrix representation of each symmetry, including the emergent ones, must also be enlarged in the bulk, with $\tilde T=\gamma_zT$, $\tilde S=S\gamma_z$, and other symmetries simply tensor product with identity matrix $I_\gamma$. One can verify that the $(d+1)$-dimensional bulk has the same symmetry as the $d$-dimensional tQCP, and its boundary has the same low energy dynamics as the $d$-dimensional tQCP of class X.

\section{Real classes}\label{sec:real}

We first discuss the eight real classes, which are further divided into two sets. The first set consists of four classes without SU(2) symmetry, namely, D, DIII, AII, and CII; while the second set consists the other four classes with SU(2) symmetry, namely, C, CI, AI, and BDI. Within each set, the emergent symmetry at the tQCP is simply the additional symmetry of the subsequent class in Table \ref{tab:Symmetry}. Connecting CII to C requires more careful analysis, which is given in Sec. \ref{sec:SU2}.  
We will use $\sigma_\mu$, $\tau_\mu$, $\eta_\mu$,  $\xi_\mu$, $\lambda_\mu$,  $\pi_\mu$, and $\zeta_\mu$ to represent Pauli matrices in different subspaces and $k_w$ and $k_u$ as the momentum of the fourth and fifth spatial dimension. We define projection operators $P_\alpha=(1+\alpha_z)/2$, where $\alpha_z$ is the Pauli matrix of one of the subspaces. We will also use subscript of the Hamiltonian to label the spatial dimension and superscript of fermion operators to label the number of its components. 

In the following examples, the Hamiltonians may have some accidental symmetries. Except for the protection symmetries listed in the last column of Table \ref{tab:Symmetry}, we allow additional perturbation terms to break these accidental symmetries but will omit these terms in the discussions.

\subsection{Classes without SU(2) symmetry}
\paragraph{D to DIII}
Let us consider a two-dimensional Hamiltonian on a square lattice
\begin{equation}
\begin{split}
    \mathcal H_2=&\sum_{\bf k}\chi_{-\bf k}^{(2)T}[\sigma_x\sin k_x+\sigma_z\sin k_y\\
    &+(2-\cos k_x-\cos k_y+m_2)\sigma_y]\chi_{\bf k}^{(2)},
\end{split}
\end{equation}
where $\chi_{\bf k}^{(2)}$ is a 2-component Majorana fermion. This Hamiltonian does not have any microscopic symmetry except for the Fermi constraint, thus belonging to class D. Physically, it describes a chiral $p$-wave superconductor \cite{Read00}.
$m_2>0$ and $-2<m_2<0$ correspond to superconductors with Chern numbers 0 and $-1$, respectively.
A topological phase transition occurs at $m_2=0$, when the energy gap closes at $\bf{k}=0$.

The tQCP can be described by a low-energy EFT
\begin{equation}
    \mathcal{H}_2=\sum_{\bf k}\chi_{-\bf k}^{(2)T}(\sigma_xk_x+\sigma_zk_y+m_2\sigma_y)\chi_{\bf k}^{(2)}.
\end{equation}
Although the EFT does not have the correct UV information, it faithfully reproduces the change of the topological invariant (Chern number) across $m_2=0$. At the tQCP $m_2=0$, there is an emergent TRS $T=i\sigma_yK$, with $T^2=-1$ in the EFT. 
The emergent TRS can always be taken as  $T=i\Gamma_0K$, where $\Gamma_0$ is the antisymmetric matrix associated with the mass term ($\Gamma_0=\sigma_y$ in this case), which anticommutes with the kinetic terms. 

If we promote the symmetry group of the tQCP $G_\text{tQCP}=Z_2^T, T^2=-1$ to the protecting symmetry, this EFT describes the low energy dynamics of the boundary of a 3D bulk Hamiltonian in class DIII with double the fermion degrees of freedom. The EFT of the bulk can be written as
\begin{equation}\label{eq:DIIIeft}
    \mathcal{H}_3=\sum_{\bf k}\chi_{-\bf k}^{(4)T}(\sigma_x\tau_zk_x+\sigma_z\tau_zk_y+\tau_xk_z+m_3\tau_y)\chi_{\bf k}^{(4)},
\end{equation}
where $\chi_{\bf k}^{(4)}$ is a 4-component Majorana fermion and $\chi_{\bf k}^{(2)}=P_\tau\chi_{\bf k}^{(4)}$ with $P_\tau=(1+\tau_z)/2$.
We can regularize the bulk EFT on the cubic lattice as
\begin{equation}\label{eq:DIII}
    \begin{split}
    \mathcal{H}_3=&\sum_{\bf k}\chi_{-\bf k}^{(4)T}[\sigma_x\tau_z\sin k_x+\sigma_z\tau_z\sin k_y+\tau_x\sin k_z\\
    &(3-\cos k_x-\cos k_y-\cos k_z+m_3)\tau_y]\chi_{\bf k}^{(4)},        
    \end{split}
\end{equation}
which is topological for $-2<m_3<0$. The TRS in the bulk has an enlarged representation $T=i\sigma_y\tau_zK$. The symmetry group of the bulk Hamiltonian is $\tilde G_p=Z_2^T$ with $T^2=-1$. Therefore, it belongs to class DIII.  Physically, this describes a time-reversal invariant $p$-wave topological superconductor \cite{Qi09}.

\paragraph{DIII to AII}
The 3D lattice Hamiltonian Eq. (\ref{eq:DIII}) has a tQCP at $m_3=0$, with $m_3>0$ and $-2<m_3<0$ corresponding to trivial and topological phases. At the tQCP, the EFT Eq. (\ref{eq:DIIIeft}) has an emergent charge U(1) symmetry $U=e^{i\theta\sigma_y\tau_x}$. The charge operator $Q=\sigma_y\tau_x$ can always be chosen as $Q=e^{i\alpha} U_T\Gamma_0$, where $\Gamma_0$ is the mass matrix in $\mathcal H_3$ ($\Gamma_0=\tau_y$ in this case) and $U_T$ is the unitary part of TRS ($U_T=i\sigma_y\tau_z$ in this case), $\alpha$ is to be chosen such that $Q$ is Hermitian. Since both $\Gamma_0$ and $U_T$ anticommute with the kinetic terms, their product $Q$ must commute with the massless Hamiltonian in the EFT. It is easy to check that  $T^{-1}U(\theta)T=U(-\theta)$. Therefore, the tQCP has an enlarged symmetry $G_\text{tQCP}=U(1)\rtimes Z_2^T$ with $T^2=-1$.

If we promote the enlarged symmetry to a protecting symmetry $\tilde G_p=U(1)\rtimes Z_2^T$, the EFT in Eq. (\ref{eq:DIIIeft}) describes the boundary of a gapped Hamiltonian of class AII in 4D with double the fermion degrees of freedom. The 4D bulk EFT is given by
\begin{equation}\label{eq:AII}
\begin{split}
    \mathcal{H}_4=\sum_{\bf k}\chi_{-\bf k}^{(8)}&(\sigma_x\tau_z\eta_zk_x+\sigma_z\tau_z\eta_zk_y+\tau_x\eta_zk_z\\
    &+\eta_xk_w+m_4\eta_y)\chi_{\bf k}^{(8)},
\end{split}
\end{equation}
where $\chi_{\bf k}^{(8)}$ is an 8-component Majorana fermion with $\chi_{\bf k}^{(4)}=P_\eta \chi_{\bf k}^{(8)}$ and $P_\eta=(1+\eta_z)/2$. 
We can regularize the 4D EFT on a hypercubic lattice
\begin{equation}\label{eq:AII}
\begin{split}
    \mathcal{H}_4=\sum_{\bf k}\chi_{-\bf k}^{(8)}&[\sigma_x\tau_z\eta_z\sin k_x+\sigma_z\tau_z\eta_z\sin k_y+\tau_x\eta_z\sin k_z\\
    &+\eta_x\sin k_w+(m_4+m'_4)\eta_y]\chi_{\bf k}^{(8)},
\end{split}
\end{equation}
with $m'_4=4-\cos k_x-\cos k_y-\cos k_z-\cos k_w$.
This Hamiltonian describes a 4-dimensional time-reversal invariant topological insulator when $-2<m_4<0$.
Clearly, the 4D bulk Hamiltonian has the same charge conservation $Q=\sigma_y\tau_x$ and an enlarged TRS $T=i\sigma_y\tau_z\eta_zK$.  The symmetry group of the 4D Hamiltonian is $\tilde G_p=U(1)\rtimes Z_2^T$ with $T^2=-1$, and the Hamiltonian therefore belongs to class AII.

A more detailed discussion of the emergent U(1) symmetry of DIII class tQCP and the effect of interactions can be found in Refs. \cite{Zhou23,Zhou24}.

\paragraph{AII to CII}
In principle, we can connect the tQCP of the 4D Hamiltonian (\ref{eq:AII}) to the boundary of a 5D bulk Hamiltonian of class CII. However, such a high spatial dimension is less relevant to physical systems. 
Instead, we examine a 2D Hamiltonian of class AII
\begin{equation}
    \begin{split}
    \mathcal{H}_2=&\sum_{\bf k}\chi_{-\bf k}^{(8)T}[\sigma_y\tau_x\xi_y\sin k_x+\tau_y\xi_y\sin k_y\\
    &+(2-\cos k_x-\cos k_y+m_2)\tau_z\xi_y]\chi_{\bf k}^{(8)},
    \end{split}
\end{equation}
which has a charge U(1) symmetry $U=e^{i\theta Q}$, $Q=\xi_y$ and a TRS $T=i\sigma_y\xi_zK$, and the protecing symmetry is $G_p=U(1)\rtimes Z_2^T$ with $T^2=-1$. $m_2>0$ and $-2<m_2<0$ correspond to trivial and topological phases, respectively.
It is more transparent to write this model using 4-component complex fermions $\psi_{\bf k}^{(4)}$,
\begin{equation}
    \begin{split}
    \mathcal{H}_2=&\sum_{\bf k} \psi_{\bf k}^{(4)}[\sigma_y\tau_x\sin k_x+\tau_y\sin k_y\\
    &+(2-\cos k_x-\cos k_y+m_2)\tau_z]\psi_{\bf k}^{(4)},    
    \end{split}
\end{equation}
which has a charge U(1) symmetry $U=e^{i\theta}$ and TRS $T=i\sigma_yK$. This can be considered as a 2D quantum spin Hall state \cite{Kane05a,Bernevig06a}.

In the following, we will focus on the Majorana basis and write the EFT for the tQCP as
\begin{equation}\label{eq:AII2}
    \mathcal{H}_2=\sum_{\bf k}\chi_{-\bf k}^{(8)T}(\sigma_y\tau_x\xi_yk_x+\tau_y\xi_yk_y+m_2\tau_z\xi_y)\chi_{\bf k}^{(8)}.
\end{equation}
At the tQCP $m_2=0$, there is an emergent PHS $C=i\sigma_y\tau_z\xi_xK$, $C^2=-1$. This PHS can always be take as $C=\Gamma_0T$, with $T$ being the TRS in $G_p$ ($T=i\sigma_y\xi_zK$ in this case) and $\Gamma_0$ the mass matrix ($\Gamma_0=\tau_z\xi_y$ in this case).
It is easy to check $TU(\theta)T^{-1}=U(-\theta)$ and $CU(\theta)C^{-1}=U(-\theta)$.
This tQCP has an enlarged symmetry $G_\text{tQCP}=U(1)\rtimes(Z_2^T\times Z_2^C)$ with $T^2=C^2=-1$.
This tQCP has the same EFT as the boundary of a 3D Hamiltonian of class CII with double the fermion degrees of freedom in the bulk. The 3D bulk EFT is given by
\begin{equation}\label{eq:CII}
\begin{split}
    \mathcal{H}_3=\sum_{\bf k}\chi_{-\bf k}^{(16)T}&(\sigma_y\tau_x\xi_y\eta_zk_x+\tau_y\xi_y\eta_zk_y\\
    &+\eta_xk_z+m_3\eta_y)\chi_{\bf k}^{(16)},
\end{split}
\end{equation}
with $\chi_{\bf k}^{(8)}=P_\eta\chi_{\bf k}^{(16)}$.
The bulk EFT can be regularized on a cubic lattice as
\begin{equation}\label{CII}
\begin{split}
    \mathcal{H}_3=&\sum_{\bf k}\chi_{-\bf k}^{(16)T}[\sigma_y\tau_x\xi_y\eta_z\sin k_x+\tau_y\xi_y\eta_z\sin k_y+\eta_x\sin k_z\\
    &+(m_3+3-\cos k_x-\cos k_y-\cos k_z)\eta_y]\chi_{\bf k}^{(16)},
\end{split}
\end{equation}
which is topological for $-2<m_3<0$.
It has charge U(1) $U(\theta)=e^{i\theta\xi_y}$, PHS $C=i\sigma_y\tau_z\xi_xK$ and an enlarged TRS $T=i\sigma_y\xi_z\eta_zK$.  The protecting symmetry group of the bulk Hamiltonian is $\tilde G_p=U(1)\rtimes(Z_2^T\times Z_2^C)$ with $T^2=-1$ and $C^2=-1$, and the Hamiltonian belongs to class CII.

\subsection{From CII to C}\label{sec:SU2}

The relation between class CII tQCPs and the boundary of class C bulk is less straightforward. As the microscopic symmetries of CII are $T$, $Q$, and $C$, while the microscopic symmetry of class C is SU(2). 
The key is to realize that at  tQCPs, class CII has two noncommutative emergent U(1) symmetries, which form an SU(2) group with the charge U(1) symmetry in the Majorana basis, and the SU(2) symmetry is not compatible with $T$ and $C$. 

Strictly speaking, for this part of discussion, we need a coset of $U(1)$ in $SU(2)$ group to emerge. As $U(1)$ is not a normal subgroup of $SU(2)$, the coset being two dimensional, i.e. 
a two-sphere $S^2$ does not form a group. But it can be constructed out of two emergent non-commuting $u(1)$ algebras and form an emergent $su(2)$ algebra along with the pre-existing 
$u(1)$ in CII class.

CII class has a microscopic charge U(1) symmetry $U=e^{i\theta\xi_y}$. In the following, we write its generator as $S_1=\xi_y$.
At $m_3=0$, the class CII EFT Eq. (\ref{eq:CII}) has an emergent U(1) symmetry $U'=e^{i\theta \sigma_y\xi_z\eta_x}$.
The generator $S_2$ for $U'$ can always be chosen as $S_2=e^{i\alpha}U_T\Gamma_0$ (in this case $U_T=i\sigma_y\xi_z\eta_z$ and $\Gamma_0=\eta_y$, $\alpha=0$). One can verify that $[S_1,S_2]=0$, as $S_1=Q=\eta_y$ must always commute with the mass matrix $\Gamma_0$ and anticommute with $U_T$.
Combining the generators of $U$ and $U'$, we find another emergent U(1) symmetry at the tQCP,  $U''=e^{i\theta \sigma_y\xi_x\eta_x}$, whose generator is $S_3=-iS_1S_2=\sigma_y\xi_x\eta_x$.
It is easy to check that $\{S_1,S_2,S_3\}$ form an su(2) algebra and the tQCP has an emergent SU(2) symmetry. We note that the generators of the SU(2) symmetry are all imaginary in the Majorana representation.
If we insist on this SU(2) symmetry, then $T=i\sigma_y\xi_z\eta_zK$ and $C=i\sigma_y\tau_z\xi_zK$ are no longer compatible with SU(2), as the SU(2) rotations do not commute with $T$ or $C$. 
By insisting on the SU(2) symmetry, we can connect the class CII tQCP to the boundary of a 4D class C bulk with double the fermion degrees of freedom. The EFT of the bulk is
\begin{equation}\label{eq:C}
\begin{split}    \mathcal{H}_4=\sum_{\bf k}\chi_{-\bf k}^{(32)T}&[(\sigma_y\tau_x\xi_y\eta_zk_x+\tau_y\xi_y\eta_zk_y+\eta_xk_z)\lambda_z\\
&+\lambda_xk_w+m_4\lambda_y]\chi_{\bf k}^{(32)},
\end{split}
\end{equation}
with $\chi_{\bf k}^{(16)}=P_\lambda\chi_{\bf k}^{(32)}$ and $P_\lambda=(1+\lambda_z)/2$. 
It can be regularized on a hypercubic lattice by substituting $k_i$ with $\sin k_i$ and adding a regularization mass $m'=4-\sum_{i=1}^4\cos k_i$.
The bulk Hamiltonian is protected by the SU(2) symmetry generated by $\{\xi_y,\sigma_y\xi_x\eta_x,\sigma_y\xi_z\eta_x\}$ and belongs to class C.

\subsection{Classes with SU(2)}
For the four classes with SU(2) symmetry, the construction procedure completely follows the other four classes without SU(2).

\paragraph{C to CI} 
In principle, the 4D class C tQCP can be connected to the boundary of a 5D class CI bulk. However, to keep our discussions relevant to physical systems, we consider the following 2-dimensional class C Hamiltonian of a superconductor on a square lattice
\begin{equation}
    \begin{split}
     \mathcal{H}_2=&\sum_{\bf k}\chi_{-\bf k}^{(8)T}[\sigma_y\tau_x\eta_y\sin k_x+\sigma_y\tau_z\eta_y\sin k_y\\
     &+(m_2+2-\cos k_x-\cos k_y)\tau_y]\chi_{\bf k}^{(8)} ,  
    \end{split}
\end{equation}
which has an SU(2) symmetry generated by $\{\sigma_x\tau_y,\sigma_y,\sigma_z\tau_y\}$ and breaks TRS. It describes a  superconductor with interband pairing or chiral $d$-wave superconductors with strong interaction \cite{Yang25}. It is topological for $-2<m_2<0$ and trivial for $m_2>0$. 
The tQCP between these two phases can be described by an EFT
\begin{equation}
    \mathcal{H}_2=\sum_{\bf k}\chi_{-\bf k}^{(8)T}(\sigma_y\tau_x\eta_yk_x+\sigma_y\tau_z\eta_yk_y+m_2\tau_y)\chi_{\bf k}^{(8)}.
\end{equation}
At $m_2=0$, it has an emergent TRS $T=i\tau_yK$, which is the combination of the mass matrix $\Gamma_0=\tau_y$ and complex conjugation $K$. The symmetry group of the tQCP is thus $G_\text{tQCP}=SU(2)\times Z_2^T$ with $T^2=-1$. It has the same IR dynamics as the boundary of a 3D bulk of class CI with double the fermion degrees of freedom. The bulk EFT can be given as
\begin{equation}\label{eq:CIeft}
\begin{split}
    \mathcal{H}_3=\sum_{\bf k}\chi_{-\bf k}^{(16)T}&[(\sigma_y\tau_x\eta_yk_x+\sigma_y\tau_z\eta_yk_y)\lambda_z\\
    &+\lambda_xk_z+m_3\lambda_y]\chi_{\bf k}^{(16)},
\end{split}
\end{equation}
where $\chi_{\bf k}^{(8)}=P_\lambda\chi_{\bf k}^{(16)}$.
We can regularize the bulk on a cubic lattice
\begin{equation}\label{eq:CI}
    \begin{split}
    \mathcal{H}_3=&\sum_{\bf k}\chi_{-\bf k}^{(16)T}[(\sigma_y\tau_x\eta_y\sin k_x+\sigma_y\tau_z\eta_y\sin k_y)\lambda_z+\lambda_x\sin k_z\\
    &+(m_3+3-\cos k_x-\cos k_y-\cos k_z)\lambda_y]\chi_{\bf k}^{(16)},    
    \end{split}
\end{equation}
which is topological for $-2<m_2<0$.
It has an enlarged TRS $T=i\tau_y\lambda_zK$ and the same SU(2) generated by $\{\sigma_x\tau_y,\sigma_y,\sigma_z\tau_y\}$. It easy to verify that SU(2) rotations commute with $T$. Hence, the bulk Hamiltonian has a protecting symmetry group $\tilde G_p=SU(2)\times Z_2^T$ with $T^2=-1$ and belongs to class CI.

We note that without requiring the Dirac form of the EFT, we can have a class C Hamiltonian with fewer fermion degrees of freedom. It represents a weakly interacting two-dimensional $d_{xy}+id_{x^2-y^2}$ superconductor, whose low energy EFT is given by 4-component Majorana fermions and has a dynamic critical exponent $z=2$. The $z=2$ tQCP of class C and the effect of interactions was analyzed in detail in Ref. \cite{Yang25}. 

\paragraph{CI to AI}
At $m_3=0$, the 3D class CI Hamiltonian (\ref{eq:CI}) is at the tQCP. Its EFT Eq. (\ref{eq:CIeft}) has an emergent charge U(1) symmetry $U=e^{i\theta Q}$, $Q=\tau_y\lambda_x$. The charge $Q$ is the product of the unitary part of TRS $U_T=i\tau_y\lambda_z$ and the mass matrix $\Gamma_0=\lambda_y$. It is easy to check that $Q$ commutes with the SU(2) generators and the tQCP has an enlarged symmetry $G_\text{tQCP}=SU(2)\times[U(1)\rtimes Z_2^T]$.
The tQCP has the same EFT as the the boundary of a 4D bulk of class AI. The EFT of the bulk can be given by 
\begin{equation}\label{eq:AI}
\begin{split}
    \mathcal{H}_4=\sum_{\bf k}\chi_{-\bf k}^{(32)}&[(\sigma_y\tau_x\eta_y\lambda_zk_x+\sigma_y\tau_z\eta_y\lambda_zk_y+\lambda_xk_z)\pi_z\\
    &+\pi_xk_w+m_4\pi_y]\chi_{\bf k}^{(32)},        
\end{split}
\end{equation}
where $\chi_{\bf k}^{(16)}=P_\pi \chi_{\bf k}^{(32)}$ and $P_\pi=(1+\pi_z)/2$. 
It can be regularized on a hypercubic lattice by substituting $k_i$ with $\sin k_i$ and adding a regularization mass $m'=4-\sum_{i=1}^4\cos k_i$.
The bulk Hamiltonian has the same charge U(1) symmetry $U=e^{i\theta\tau_y\lambda_x}$, spin SU(2) symmetry generated by $\{\sigma_x\tau_y,\sigma_y,\sigma_z\tau_y\}$, and an enlarged TRS $T=i\tau_y\lambda_z\pi_zK$. It is easy to verify that SU(2) rotations commute with $T$ but charge U(1) does not commute with $T$. Therefore, the bulk Hamiltonian has the protecting symmetry group $\tilde G_p=SU(2)\times[U(1)\rtimes Z_2^T]$ with $T^2=-1$ and belongs to class AI.

\paragraph{AI to BDI} 
Class AI is topologically trivial in $d=1,2,3$, and a $d=0$ critical state is simply a topological zero mode protected by symmetries. Therefore, we will consider the tQCP of the $d=4$ Hamiltonian, whose EFT is given by Eq. (\ref{eq:AI}). At $m_4=0$, the tQCP has an emergent PHS $C=i\tau_y\lambda_z\pi_xK$ with $C^2=-1$. The PHS is the product of the mass matrix $\Gamma_0=\pi_y$ and TRS $T=i\tau_y\lambda_z\pi_zK$. 
The 4D tQCP has an enlarged symmetry $G_\text{tQCP}=SU(2)\times[U(1)\rtimes(Z_2^T\times Z_2^C)]$, with $T^2=C^2=-1$. It has the same EFT as the boundary of a 5D bulk of class BDI with double the fermion degrees of freedom. The bulk EFT is given by
\begin{equation}
\begin{split}
    \mathcal{H}_5=\sum_{\bf k}\chi_{-\bf k}^{(64)T}[(\sigma_y\tau_x\eta_y\lambda_z\pi_zk_x+\sigma_y\tau_z\eta_y\lambda_z\pi_zk_y\\
    +\lambda_x\pi_zk_z+\pi_xk_w)\zeta_z+\zeta_xk_u+m_5\zeta_y]\chi_{\bf k}^{(64)}
\end{split}    
\end{equation}
where $\chi_{\bf k}^{(32)}=P_\zeta\chi_{\bf }k^{(64)}$ and $P_\zeta=(1+\zeta_z)/2$. It can be regularized on a hypercubic lattice by substituting $k_i$ by $\sin k_i$ and adding a regularization mass $m'=5-\sum_{i=1}^5\cos k_i$.
The bulk Hamiltonian has a charge U(1) symmetry $U(\theta)=e^{i\theta \tau_y\lambda_x}$ and spin SU(2) symmetry generated by $\{\sigma_x\tau_y,\sigma_y,\sigma_z\tau_y\}$, PHS $C=i\tau_y\lambda_z\pi_xK$, and an enlarged TRS $T=i\tau_y\lambda_z\pi_z\zeta_zK$. 
The bulk Hamiltonian has the symmetry group $SU(2)\times[U(1)\rtimes(Z_2^T\times Z_2^C)]$ with $T^2=C^2=-1$ and belongs to class BDI.

\subsection{From BDI to D}\label{sec:closure}
To complete the Bott clock, we finally consider connecting BDI tQCPs to the boundary of class D bulk. To keep our discussions relevant to physical systems, we will again lower the physical dimension and consider a class BDI Hamiltonian in $d=1$.
However, the boundary states of class D superconductors in 2D are always chiral but the 1D tQCP is nonchiral. Therefore, we are not able to construct a holographic theory between class BDI tQCPs in 1D and gapped class D bulk in 2D.
Nevertheless, we can construct a 2D bulk using emergent symmetries. We leave such discussions to the Appendix.

\section{Complex classes}\label{sec:complex}
Next, we present the holographic theory of tQCPs for complex classes.  
\paragraph{A to AIII}
Let us start with a 2D lattice model of class A, which is the familiar Qi-Wu-Zhang model \cite{QWZ},
\begin{equation}
\begin{split}
    \mathcal{H}_2=&\sum_{\bf k}\psi_{\bf k}^{(2)\dagger}[\sigma_x\sin k_x +\sigma_z\sin k_y\\
    &+(2-\cos  k_x-\cos k_y+m_2)\sigma_y ]\psi_{\bf k}^{(2)},
\end{split}    
\end{equation}
where $\psi_{\bf k}^{(2)}$ is a 2-component complex fermion. 
It has Chern number 0 and $-1$ for $m_2>0$ and $-2<m_2<0$, respectively.
We can write the EFT for the tQCP at $m_2=0$ as
\begin{equation}\label{eq:A}
    \mathcal{H}_2=\sum_{\bf k}\psi_{\bf k}^{(2)\dagger}(\sigma_x k_x+\sigma_z k_y+\sigma_y m_2)\psi_{\bf k}^{(2)},
\end{equation}
At $m_2=0$, it acquire an emergent sublattice symmetry $S=\sigma_y$. Since the mass matrix $\Gamma_0$ always anticommutes with the kinetic terms, one can always choose $S=\Gamma_0$. 
We can construct a 3D bulk Hamiltonian of class AIII with double the band degrees of freedom whose boundary has the same EFT as the tQCP. The bulk EFT is given by
\begin{equation}
    \mathcal{H}_3 =\sum_{\bf k}\psi_{\bf k}^{(4)\dagger}(\sigma_x\tau_zk_x+\sigma_z\tau_zk_y+\tau_xk_z+m_3\tau_y)\psi_{\bf k}^{(4)},    
\end{equation}
where $\psi_{\bf k}^{(4)}$ is a 4-component complex fermion with $\psi_{\bf k}^{(2)}=P_\tau\psi_{\bf k}^{(4)}$ and $P_\tau=(1+\tau_z)/2$. 
It can be regularized on a square lattice as
\begin{equation}\label{eq:AIII}
    \begin{split}
            \mathcal{H}_3 =&\sum_{\bf k}\psi_{\bf k}^{(4)\dagger}[\sigma_x\tau_z\sin k_x+\sigma_z\tau_z\sin k_y+\tau_x\sin k_z\\
            &+(3-\cos k_x-\cos k_y-\cos k_z+m_3)\tau_y]\psi_{\bf k}^{(4)}, 
    \end{split}
\end{equation}
which is topological for $-2<m_3<0$. The 3D bulk has an enlarged sublattice symmetry $S=\sigma_y\tau_z$, thus belonging to class AIII.

In the Majorana basis, the EFT of class A Eq. (\ref{eq:A}) can be written in momentum space as 
\begin{equation}
    \mathcal{H}_2=\sum_{\bf k}\chi_{-\bf k}^{(4)T}(\sigma_x k_x+\sigma_z k_y+\sigma_y m_2)\otimes I_2\chi_{\bf k}^{(4)},
\end{equation}
with $I_2$ a two-by-two identity matrix in the Majorana $\{+,-\}$ subspace. Charge conservation is manifested as a pseduospin $U(1)$ symmetry $U=e^{i\theta s_y}$, where $s_y$ is the antisymmetric Pauli matrix in the same $\{+,-\}$ Majorana subspace. At $m_2=0$, it acquires an emergent TRS $T=i\sigma_yK$. The emergent TRS can always be chosen as $T=i\Gamma_0K$, as $\Gamma_0$ is antisymmetric in the Majorana basis and anticommutes with the kinetic terms. We note that the TRS and pseduospin U(1) commute. Thus, the tQCP has an enlarged symmetry $G_\text{tQCP}=U(1)\times Z_2^T$.
The corresponding 3D bulk EFT is
\begin{equation}
    \mathcal{H}_3=\sum_{\bf k}\chi_{-\bf k}^{(8)T}(\sigma_x\tau_z k_x+\sigma_z\tau_z k_y+\tau_xk_z+\tau_y m_3)\otimes I_2\chi_{\bf k}^{(8)},
\end{equation}
where $\chi_{\bf k}^{(8)}$ is an 8-component Majorana fermion with $\chi_{\bf k}^{(4)}=P_\tau\chi_{\bf k}^{(8)}$. It can be regularized on a cubic lattice as
\begin{equation}
    \begin{split}
            \mathcal{H}_3 =&\sum_{\bf k}\chi_{-\bf k}^{(8)T}[\sigma_x\tau_z\sin k_x+\sigma_z\tau_z\sin k_y+\tau_x\sin k_z\\
            &+(3-\cos k_x-\cos k_y-\cos k_z+m_3)\tau_y]\otimes I_2\chi_{\bf k}^{(8)}, 
    \end{split}
\end{equation}
which is topological for $-2<m_2<0$. The 3D bulk has the same pseduospin U(1) symmetry and an enlarged TRS $T=i\sigma_y\tau_zK$ that commute with each other. The protecting symmetry group of the bulk Hamiltonian is $\tilde G_p=U(1)\times Z_2^T$, realizing the AIII class. When converted back to the complex fermion basis, the Hamiltonian is identical to Eq. (\ref{eq:AIII}).

\paragraph{AIII to A}
The boundary of class A states are chiral while the AIII tQCPs are nonchiral. Therefore, we cannot construct a holographic theory of class AIII tQCP and a class A bulk. Nevertheless, we can construct a bulk theory for the tQCP using emergent symmetries. We leave this to the Appendix.

\section{Discussions}\label{sec:discussion}
We note that the emergent symmetry at the tQCP is not unique.
This means a given tQCP can be connected to the boundaries of different bulk Hamiltonians of different symmetry classes. 
However, we can always choose the emergent symmetry to be the additional symmetry in the subsequent class in the Bott clock. In this case, the $d$-dimensional tQCP and the $(d+1)$-dimensional bulk have the same classification.
For example, we have explicitly constructed a 3D bulk Hamiltonian of class AIII, whose boundary has the same IR dynamics as a 2D tQCP of class A (see Sev. \ref{sec:complex}).
But one can check that in the complex fermion representation the emergent symmetry of Eq. (\ref{eq:A}) at $m_2=0$ can also be a TRS $T=i\sigma_yK$ with $T^2=-1$. In this complex fermion representation, the charge U(1) symmetry is given by $U=e^{i\theta}$. If we insist on this TRS as the emergent symmetry, we will construct a 3D bulk with a protecting symmetry group $\tilde G_p=U(1)\rtimes Z_2^T$ with $T^2=-1$, which belongs to class AII. However, the topological classification of the bulk and boundary will be different in this case. Class AII topological insulators are classified by a $Z_2$ index in 3D \cite{Fu07}, while class A Chern insulators and its tQCPs are classified by an integer in 2D. Therefore, with this alternative construction, the holographic theory will break down beyond the minimal model.

The gapless boundaries of SPTs are known to be robust, as a direct manifestation of 't Hooft anomalies.
In this article on the other hand, we have fully utilized the idea of emergent symmetry at tQCPs to carry out discussions of holographic theories for tQCPs.
These studies strongly suggest that the anomalous emergent symmetries at tQCPs can be intimately related to 't Hooft anomalies in the gapless boundaries of SPTs.

There still remains an important question: what is the proper topological EFT for such an anomalous emergent symmetry?
One of the ways to take care of the proper ultraviolet physics in an infrared EFT is to introduce a topological term that takes care of the anomalous symmetry flow and to saturate the anomaly, similar to the better known inflow from the bulk to a boundary. Recent efforts along this direction for non-invertible topologically ordered phases indicate
the existence of conformal manifolds once inflow is taken into account \cite{Saran25,Saran26,Tianyao26}. 
This remains to be done explicitly and systematically for invertible SPTs studied here. In topological superconductors, special care is needed to pay attention to gravitational anomalies (mixed). We plan to study this explicitly in the future.

An equally fascinating question is about if all emergent 
symmetry at tQCPs around the Bott clock are actually approximate symmetries of much larger non-local symmetry structures. At specific tQCPs in DIII class SPTs, it has been illustrated that emergent symmetries exhibit complicated non-on site non-compact structures\cite{Zhou26}, similar to those found in the studies of lattice chiral fermions for field theoretical models\cite{Shao25,Thorngren26}. Around the Bott clock, how the non-local emergent symmetry groups are directly related to the exact protecting symmetry $G_p$ and other topological data remains an open question. These questions must be properly addressed before we can fully understand IR emergent symmetries at tQCPs.

\begin{acknowledgments}
We thank Wei-Cheng Ye for helpful discussions during the early stage of this project. F.Y. is supported by the National Natural Science Foundation of China (Grand No. 12504187) and the Start-up Research Fund of Southeast University (RF1028624190). F.Z. is in part supported by an NSERC(Canada) Discovery grant under contract number RGPIN-2020-07070.

\end{acknowledgments}

\appendix

\section{BDI class}
If we take the tenfold-way language and allow spinless fermions, the minimal BDI model is a single Kitaev chain with imposed spinless TRS $T=\sigma_zK$ and Fermi constraint viewed as PHS $C=K$,
\begin{equation}
    \mathcal{H}_1=\sum_{\bf k}\chi_{-\bf k}^{(2)T}[\sigma_x\sin k_x+(m_1+1-\cos k)\sigma_y]\chi_{\bf k}^{(2)},
\end{equation}
which is trivial for $m_1>0$ and topological for $-2<m_1<0$.
The EFT at the tQCP is given by
\begin{equation}\label{eq:1BDI}
    \mathcal{H}_1=\sum_{\bf k}\chi_{-\bf k}^{(2)T}(\sigma_xk_x+m_1\sigma_y)\chi_{\bf k}^{(2)}.
\end{equation}
At the tQCP $m_1=0$, it has an emergent spin-1/2 TRS symmetry $T'=i\sigma_yK$.

By keeping these symmetries, we can construct a 2D bulk EFT with double the fermion degrees of freedom
\begin{equation}
    \mathcal{H}_2=\sum_{\bf k}\chi_{-\bf k}^{(4)T}(\sigma_x\tau_zk_x+\tau_xk_y+m_2\tau_y)\chi_{\bf k}^{(4)}
\end{equation}
where $\chi_{\bf k}^{(2)}=P_\tau\chi_{\bf k}^{(4)}$. 
It can be regularized on a square lattice
\begin{equation}
\begin{split}    \mathcal{H}_2=&\sum_{\bf k}\chi_{-\bf k}^{(4)T}[\sigma_x\tau_z\sin k_x+\tau_x\sin k_y\\
&+(m_2+2-\cos k_x-\cos k_y)\tau_y]\chi_{\bf k}^{(4)},
\end{split}
\end{equation}
which is topological for $-2<m_2<0$.
In the tenfold way language, the bulk Hamiltonian matrix has the same PHS $C=K$, and two enlarged TRS, one spinless $T=\sigma_z\tau_zK$ and the other spin-1/2  $T'=i\sigma_y\tau_zK$.
We note that this bulk Hamiltonian should not be interpreted as the class DIII $p$-wave superconductor, as we require an extra spinless TRS $T=\sigma_z\tau_zK$. 
These two TRS's combine to a unitary symmetry $\sigma_x$. We can therefore block diagonalize the bulk Hamiltonian by $\sigma_x$, and each block belongs to class D.

\section{AIII class}
Let us consider a 1D lattice model of class AIII, which is the familiar Su-Shrieffer-Heeger model \cite{SSH}
\begin{equation}
    \mathcal{H}_1=\sum_{\bf k}\psi_{\bf k}^{(2)\dagger}[\sigma_x\sin k_x+(m_1+1-\cos k)\sigma_y]\psi_{\bf k}^{(2)},
\end{equation}
with sublattice symmetry $S=\sigma_z$ and charge U(1) $U=e^{i\theta}$. $m_2>0$ and $-2<m_2<0$ correspond to trivial and topological phases, respectively.
The EFT 
\begin{equation}
    \mathcal{H}_1=\sum_{\bf k}\psi_{\bf k}^{(2)\dagger}(\sigma_x  k_x+m_1 \sigma_y)\psi_{\bf k}^{(2)},
\end{equation}
has another emergent sublattice symmetry $S'=\sigma_y$ at tQCP $m_2=0$. We can construct a bulk EFT with charge U(1) and both of the two enlarged sublattice symmetries $S=\sigma_z\tau_z$ and $S'=\sigma_y\tau_z$
\begin{equation}
    \mathcal{H}_2=\sum_{\bf k}\psi_{\bf k}^{(4)\dagger}(\sigma_x\tau_zk_x+\tau_xk_y+m_2\tau_y)\psi_{\bf k}^{(4)}.
\end{equation}
The bulk EFT can be regularized on a square lattice 
\begin{equation}
\begin{split}    \mathcal{H}_2=&\sum_{\bf k}\psi_{\bf k}^{(4)\dagger}[\sigma_x\tau_z\sin k_x+\tau_x\sin k_y\\
&+(m_2+2-\cos k_x-\cos k_y)\tau_y]\psi_{\bf k}^{(4)},
\end{split}
\end{equation}
which is topological for $-2<m_2<0$.
We note that this should not be interpreted as a topological insulator of class AII as we require the protecting symmetry to be two anticommuting sublattice symmetries rather than a spin-1/2 TRS. These two sublattice symmetries combine to a unitary symmetry $\sigma_x$. We can block diagonalize the bulk Hamiltonian with $\sigma_x$ and each block belongs to class A. 

\bibliography{references}

\end{document}